\documentclass[12pt]{article}

\usepackage{scicite}

\usepackage{mathptmx}
\usepackage{graphicx,cite}
\usepackage{times}
\graphicspath{{./imgs/}}
\usepackage{hyperref}
\usepackage{amsmath}
\usepackage{amssymb}
\usepackage{booktabs}
\usepackage{url}
\usepackage{breakurl}
\usepackage{subfigure}
\usepackage{hyperref}
\usepackage{gensymb}
\usepackage[final]{microtype}
\usepackage[T1]{fontenc}
\usepackage{soul}
\usepackage{color}

\usepackage{times}

\newenvironment{sciabstract}{%
\begin{quote} \bf}
{\end{quote}}

\newcounter{lastnote}

\title{Google Cardboard Dates Augmented Reality : Issues, Challenges and Future Opportunities}

\author
{Ramakrishna Perla and Ramya Hebbalaguppe\\
\\
\normalsize{TCS Research, New Delhi, India}\\
\\
\normalsize{E-mail: r.perla@tcs.com, ramya.hebbalaguppe@tcs.com}
}

\date{}

\begin{document} 


\baselineskip24pt


\maketitle 

\section*{Abstract}
\begin{sciabstract}
%\abstractname{}
 The Google's frugal Cardboard solution for immersive Virtual Reality (VR) experiences has come a long way in the VR market. The Google Cardboard VR applications will support us in the fields such as education, virtual tourism, entertainment, gaming, design etc. Recently, Qualcomm's Vuforia SDK has introduced support for developing mixed reality applications for Google Cardboard which can combine Virtual and Augmented Reality (AR) to develop exciting and immersive experiences. In this work, we present a comprehensive review of Google Cardboard for AR and also highlight its technical and subjective limitations by conducting a feasibility study through the inspection of a Desktop computer use-case. Additionally, we recommend the future avenues for the Google Cardboard in AR. This work also serves as a guide for Android/iOS developers as there are no published scholarly articles or well documented studies exclusively on Google Cardboard with both user and developer's experience captured at one place.
\end{sciabstract}

% In setting up this template for *Science* papers, we've used both
% the \section* command and the \paragraph* command for topical
% divisions.  Which you use will of course depend on the type of paper
% you're writing.  Review Articles tend to have displayed headings, for
% which \section* is more appropriate; Research Articles, when they have
% formal topical divisions at all, tend to signal them with bold text
% that runs into the paragraph, for which \paragraph* is the right
% choice.  Either way, use the asterisk (*) modifier, as shown, to

\section{Introduction}
\label{intro}

Recent advances in hardware and software mobile technology have enabled ways for developments in mobile VR/AR applications. Google Cardboard is a VR and AR platform developed by Google for the use with head-mount for a smartphone  \ref{branstetter}\ref{pierce}.  To date, the current widespread Cardboard VR applications have shown what Google Cardboard is today. Google announced that in the platform's first 19 months, over 5 million Cardboard viewers had shipped, over 1,000 compatible applications had been published, and over 25 million application installs had been made\ref{google_blog}.

%However, the advances of mobile VR and AR technology has met with its own limitations, which resulted in not able to reach mass-market~\cite{nazri2014current} (See Section \ref{sec:limGC}). 

\begin{figure}[ht!]
	\begin{center}
		\includegraphics[width=0.31\textwidth]{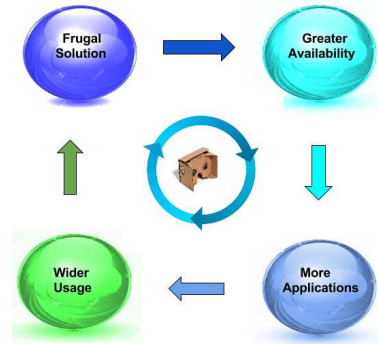} % Method_Flow_2.pdf
	\end{center}
	\caption{The parameters fueling the growth of Google Cardboard based VR and AR applications.}
\label{cb_cycle}
\end{figure}

%definition
Google Cardboard is launched at Google I/O 2014 developers conference to encourage development of VR and AR Android applications, with a release to iOS at the following year's event \ref{branstetter}\ref{pierce}. Many high cost sophisticated VR headsets exist in the market but Google's intent was to bring VR to mass-market through frugal Cardboard design. Figure \ref{cb_cycle} shows the factors making Google cardboard's reach viable for VR and AR.

%Design of Cardboard was motivated for applicati (??ons such as ... 

\emph{Design of Google Cardboard and its evolution}: A Cardboard is designed using simple stiff cardboard and anyone can either construct their own headset using simple materials such as velcro, tape adhesive, two $45mm$ focal length lenses, two magnets and most importantly, simple stiff cardboard using the specifications published by Google or purchase a assembled kit from any online vendors. A smartphone can be inserted into the headset to experience the Google Cardboard compatible applications.

\begin{figure}[ht!]
\begin{center}
   \includegraphics[width=0.8\linewidth]{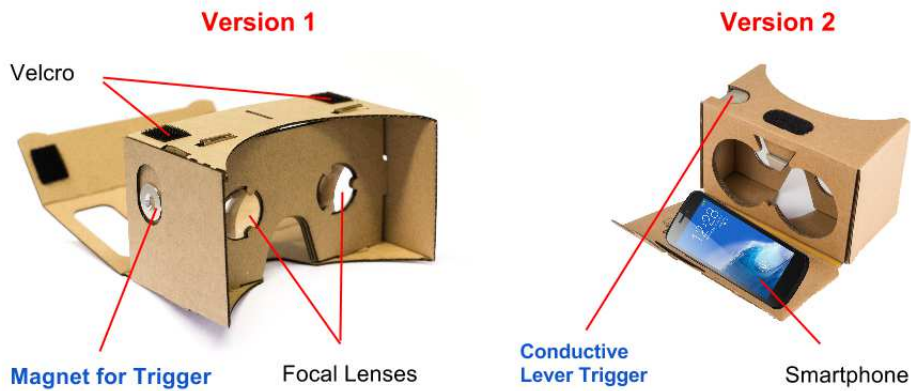}
\end{center}
   \caption{Version 1 of the Cardboard sports a magnetic trigger while version 2 of Cardboard utilizes a conductive lever for click event. Refer Introduction for problems pertaining to magnetic trigger. }
\label{fig:cardDes}
\end{figure}

%operation and software functionality
There are two versions of Google Cardboard design specifications, the initial design supports phones with screens up to $5.7$ inches (\emph{140 mm}). The magnetic button located on one side of Cardboard is used as a trigger to create a touch event in the Cardboard application (see Figure \ref{fig:cardDes}). Most of the low-end smartphones without compass would not be able to utilize the magnetic button to trigger. We also observed that due to the lack of magnetic strength and smaller range of movement within the slot, the \emph{magnetic switch} could not always be used as a trigger while conducting experiments. The second version of Cardboard design supports screens up to 6 inches (\emph{150 mm}) and most importantly, the magnetic button which was found troublesome on the first official template is taken off. This is replaced by a conductive lever button (see Figure \ref{fig:cardDes}) which triggers a lever inside the headset that comes down and presses the screen to register this touch as an event to perform any application specific pre-defined task~\ref{shanklin}.
%operation and software functionality
The Cardboard compatible applications partition the smartphone display into two parts, while also applying barrel distortion, resulting in a stereoscopic (3D) vision with a wide field of view~\ref{unity}.

%The Cardboard compatible applications partition the smartphone display into two parts (Note that the Cardboard applications work in device landscape orientation), one for each eye, while also applying barrel distortion. The result is a stereoscopic (3D) vision with a wide field of view~\cite{unity-ref-gc}. Google also released two software development kits(SDKs) for developing applications using OpenGL\cite{opengl} graphics application programming interface (API): one for Android using \emph{Java} programming language and one for the game engine Unity using \emph{$C\#$} programming language.

%%other similar VR headset manufactures 

%This cheapest open design of Google Cardboard to experience full immersion VR has drawn lot of attention by both manufactures and developers lately. Now, there are several manufactures who are making similar headsets in different shapes and sizes inspired by Google Cardboard instructions. Google reported, through January 2016, over 5 million Cardboard viewers had shipped and over 1,000 compatible applications had been published~\cite{gargantini2015low}.% 
Google also encourages other manufactures in designing VR viewers by allowing them to generate their own Viewer Profile (include fields such as Screen-to-Lens distance, Inter-Lens distance, Distortion coefficients and Field-of-View angles) to make sure that their viewer will work seamlessly with Google Cardboard applications~\ref{vr_profile}. This avoids the problems relating to stereoscopic vision such as double images or blurry picture observed when the viewer specifications deviate or slightly vary from Cardboard viewer specifications. We infer from \ref{hypergrid} that (a) Google Cardboard is the most economical gear for VR/AR applications, (b) Cardboard allows developments in AR through aperture in front-pad enabling the video-see-through mode, (c) Supports all the screen sizes upto 6 inches, and (d)  Low weight among all the VR/AR headsets. This motivated the use of Google Cardboard different AR/VR applications. The recent work by Perla et al. \ref{perla_inspection} has discussed an industrial inspection framework using multiple AR devices where extension of Google Cardboard, which was initially envisioned for VR, was also extended to AR. Further, Hegde et al. \ref{hegde_gestures} proposed simple hand swipe gestures using GMM based hand modeling of skin pixels data in egocentric view to navigate options on wearable device. This work is extended to recognize more robust hand swipe gestures in \ref{mohatta_gestures}, and the results are compared against existing methods. In \ref{gupta_roi}, Gupta et al. presented the idea of highlighting the ROI for frugal AR headsets.

We summarise the key contributions of our paper:

\begin{itemize}
\item[1] We provide a comprehensive review of Google Cardboard focusing on its design and evolution for AR (Refer Introduction \ref{intro}).
\item[2]  A sample AR application using Google Cardboard has been developed for inspection assistance of a Desktop computer. Using this as a platform, a feasibility study was conducted to explore Cardboard for AR. We captured set of subjective metrics in addition to the suggestions on technical/hardware limitations of developing AR applications for Cardboard (Refer Sections \ref{sec:limGC} and \ref{sec:exps}). 
\item[3] We also suggest future research directions and opportunities for developing AR applications on Google Cardboard in various domains (Refer Section \ref{sec:discussion} and \ref{sec:futOpp}). There has been no existing scholarly articles addressing this. 
\end{itemize}
%
%The organisation of the paper is as follows: In Section \ref{sec:ar_overview}, an overview of AR for a generic AR system is discussed. Section \ref{sec:limGC} describes the technical and non-technical limitations of using Google Cardboard for AR. Discussion and future opportunities are discussed in Sections \ref{sec:discussion} and \ref{sec:futOpp}. Finally, we conclude with the findings that could facilitate the use of Google Cardboard for AR despite its limitations. 

\begin{figure*}[ht!]
\begin{center}
   \includegraphics[width=0.8\linewidth]{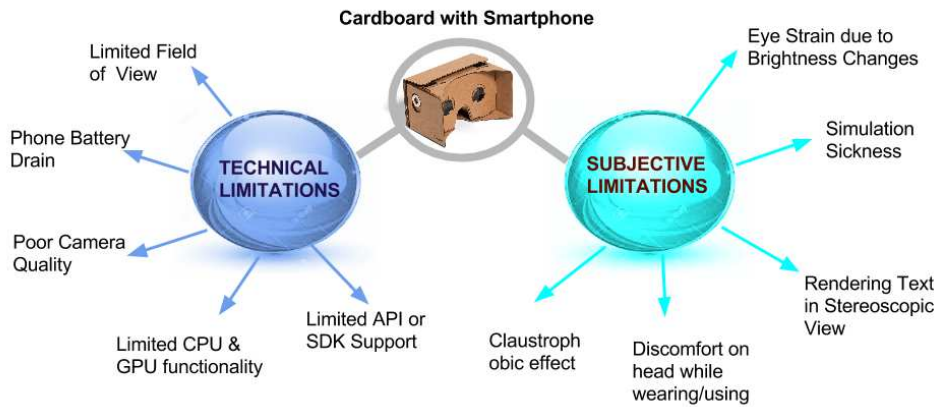}
\end{center}
   \caption{This shows the Technical and Subjective limitations of using Cardboard for AR. }
\label{fig:GCMainFig}
\end{figure*}

\section{Limitations of Google Cardboard for AR}
\label{sec:limGC}

The technological demands for AR are much higher than for VR, which is why the field of AR
took longer to mature than that of VR \ref{van}. However, the key components, namely displays, trackers, graphics computers, and software needed to build an AR system have remained the same since Ivan Sutherland's pioneering work of the 1960s \ref{sutherland}. As the smartphone technology is growing at a rapid rate, there is an increasing possibility of producing accurate and efficient key components of AR using just a smartphone. This motivates developers to design intuitive interfaces, and researchers to work on user-friendly interactions for Google Cardboard.

To test the feasibility of Google Cardboard for AR, we developed a sample AR application for the use with Google Cardboard for the inspection of a Desktop computer. The sample inspection process consisted of following steps: 
\begin{itemize}
\item[1] Reading part specific inspection data from the files stored on remote server,
\item[2] Questions are loaded into the application and sequentially overlaid on to the device screen contextually,
\item[3] Google's speech-to-text is integrated with the application; this service continuously listens for the speech input,
\item[4] Once the particular inspection item is addressed (users are instructed to say either "YES" or "NO" or "SKIP"), the data is stored in a file and sent to the server upon the completion of inspection. 
\end{itemize}

The feasibility study through post-inspection questionnaire is performed by considering 20 subjects and results are reported in Section \ref{sec:exps}.

Despite the opportunities to develop AR applications using Google Cardboard, there are few commercial mobile AR use-cases which cannot be implemented for the use with Google Cardboard. This is due to its limitations in both technical and non-technical aspects from developer and user perspectives. Figure \ref{fig:GCMainFig} highlights some key limitations and challenges of Google Cardboard for AR:
%In this section, we mostly concentrate on inspection or maintenance task aiding related AR applications development on Android/iOS devices for the use with Google Cardboard. more comprehensive overviews of AR concepts, technologies, and techniques can be found elsewhere. Displays, trackers, and AR systems in general need to become more
%accurate, lighter, cheaper, and less power consuming\cite{azuma2001recent} \cite{azuma1997survey}.

\subsection{Technical Limitations and Challenges}

\subsubsection{Field of View}
\label{subsec:FoV}

Google Cardboard offers limited FoV $(\approx90^{\circ})$, which is very constrained especially when developing applications for AR where user has to work outdoors; the user can never make an eye contact with the people surrounding the user and may not be aware of their presence. This could prove to be dangerous in industrial inspection when we deal with hot electronic and mechanical parts and also when the parts are in motion, this may cause damage to the inspector.
% have to be inserted somewhere else - Cardboard comes under the video-see-through category of HMDs. 

\subsubsection{User Interaction}
\label{subsec:UI}
The user interaction using Cardboard with smartphone can be in two ways: 
\begin{itemize}
\item \textit{Head Rotation}: Head rotation can be used for user interaction by placing the UI controls in the user FoV.  It requires continuously updating the UI controls location in the current FoV for the applications allowing movement of the user. If the controls are placed out of current FoV, by fixing them in one specific direction based on device compass readings, users will often need to wait, sense confusion, and have to look around for the controls \ref{phyConsiderations}. But for the applications such as inspection requiring user input at every step and continuous movement of the user head in $360^{\circ}$, using head rotation based interaction shifts the focus from the object being inspected and hence cannot be used in equipment inspections that need focus and attention for extended period of time.
\item \textit{Speech-to-Text based}: Triggering an event is done by displaying relevant speech keywords to be spoken by the user at each instance or by asking the user to go through the list of keywords at the launch of application. A voice based interaction on Android devices is achieved through the Google speech recognition API service \ref{speech}. Speech based interaction can perform better for the use of quiet indoor applications as it is difficult for the recognizer to detect speech keywords in a noisy environment.
\end{itemize}

\subsubsection{Smartphone Battery Drain}
\label{subsec:batt_Drain}

Another important consideration when developing applications on mobile platforms is battery life of smartphone. As the Cardboard AR applications require continuous camera feed in the background to view the physical world, it consume significant amount of the device battery. If the application uses speech based user interaction, the device microphone is in continuous listen mode for user speech input; this drains the device battery rapidly. Phone might heat up causing uneasiness to the user as the phone radiates the heat around the user's face.

\subsubsection{Camera Quality}
\label{subsec:CamQuality}
Since Google Cardboard is a video-see-through device, camera quality of the device is very important to view the physical world. A low end device with poor image acquisition will have a problem; Poor picture quality in low lighting condition poses problems to the users under limited illumination scenarios. If there exists  digital/electronic display in the user FoV, camera reflections impair the visibility. 
%
%-camera reflections when used in very high illumination condition\\
%	-poor picture quality when used in low lighting conditions\\
%	-requires very good camera quality phone with better processing power as AR apps implements graphics rendering in the background\\
%	
\subsubsection{Limited API and SDK Support}
\label{limtedAPI}
Applications need SDK or API support to implement tasks such as lens distortion correction, 3D calibration, side-by-side rendering, and stereo geometry configuration. While Google and Qualcomm have SDK's providing support for Cardboard, only Qualcomm SDK supports AR through Vuforia SDK \ref{vuforia}. Cardboard SDK \ref{cardboard_sdk} is developed for VR, and it uses Open GL graphics API in the background. Since there are no specific API's to achieve AR experience using Cardboard SDK, we overcome this by drawing camera into Open GL texture and display it using Cardboard-view \emph{stereorenderer} methods.
\subsubsection{Other Hardware Limitations}
\label{subsec:OtherhardwareLimitations}
Other hardware limitations include low processing power and internal memory, outdated processors, low-quality displays, poor camera quality and sensor imperfections are some of the issues \ref{technavio}. The device requires an implementation of Open GL graphics pipeline provided by the device manufacturer \ref{opengl}; low-end devices may not be capable of handling graphics rendering. There are number of mobile resources which are required to be used in parallel for AR application such as:
\begin{itemize}
\setlength\itemsep{0em}
\item Camera feed is always ON in order to show the physical world which is to be augmented; this consumes significant amount of battery when used for longer durations, 
\item Microphone always listening to voice input and sometimes noise and keeps comparing with pre-defined set of keywords drains battery. Use of head rotation based user interaction may be helpful for certain applications requiring less user movement and input,
\item Depends on internet connectivity as speech data is always being sent to Google servers. Usage of offline speech recognition library such as Pocketsphinx may not be a good solution as the \emph{speech-to-text} fidelity is no where comparable to the Google speech engine which is trained over huge corpus,
\item Sending live camera stream or sensors data to remote server to understand the environment for accurate overlay of virtual data. This may be required as there is limited API/library support for running computer vision, machine learning, or deep learning algorithms on smartphone,
\item Displaying virtual graphic overlays (3D graphics rendering).
\end{itemize}
  Given these requirements, application runs smartphone CPU and GPU both at maximum, thereby heating up the device quickly.
	
\subsection{Subjective Limitations}

There are some important physiological considerations in designing applications for Google Cardboard that can restrict the user if not handled well \ref{phyConsiderations}. %The most important issues limiting the wider use of Google Cardboard are as follows:

\subsubsection{Simulation Sickness}
\label{subsubsec:simSickness}

Simulation sickness is the result of a  disparity between a perceived experience and what one actually experiences  \ref{phyConsiderations}. This is a big challenge for VR/AR Cardboard applications. Manufacturers are working on designing headsets with better optics and advanced head tracking technology that could solve this problem in the future to enable users to comfortably use the headsets for longer duration \ref{technavio}.
%\footnote{http://www.technavio.com/blog/the-pros-and-cons-of-google-cardboard#sthash.o2YVHp0h.dpuf}. 
%
%Low-cost? Yes. Accessible? Yes. A perfect solution? Not so much. Simulation sickness—which is the result of a disparity between a perceived experience and what one actually experiences— is a big challenge for the Google Cardboard.
%
%Vendors are working on developing better optics with advanced head tracking technology that could solve this problem in the next few years, in the hopes that users will be able to use the device for more than 15 minutes at a time.
%
%- See more at: http://www.technavio.com/blog/the-pros-and-cons-of-google-cardboard#sthash.o2YVHp0h.dpuf

\subsubsection{Rendering text in Stereoscopic view}
\label{subsubsec:steroRen}	
	%When developing applications for Google Cardboard, user is completely immersed in virual/augmented scene shown in the stereoscopic view on the device screen.%
For VR applications, rendering text or any other graphics can be done by  binding it to a pre-defined location on the virtual scene containing some different colored area (as it helps user in easy reading) and tracking the head rotation to avoid nausea \ref{phyConsiderations}.  But for AR applications having no reference from physical world co-ordinate system to camera co-ordinate system, overlaying the text at a fixed location on the screen display for longer duration may cause discomfort/eye-strain. The reason being that the overlaid-text is independent of the background physical world scene which is continuously changing according to the user movement. 
%The reason for this is that, even though the head rotation tracking is possible, we don't have any prior knowledge of what is coming in the user FoV to bind the text instructions in specific position on the scene.  
	
% -Render 2D splash screens in 3D space\\	
%	Fixing the splash screen, or any graphic, to the user’s head and turning off head tracking may cause discomfort[x7].
%
%When displaying a splash screen, with a logo or title sequence, render 2D sprites in a 3D virtual space and maintain head tracking. VR applications using splash screens that only track head rotation to one degree of freedom (1DOF) can avoid nausea for most users, but 3DOF (rotation, pitch, and yaw) is still preferable.

\subsubsection{Brightness of Virtual Content}

%Another important consideration while developing AR apps for Cardboard is 
Change in brightness of the virtual content overlaid is an issue given the limited screen size and limited FoV; the immediate transition from the dark scene to a bright scene causes eye strain and restricts the user for longer duration tasks \ref{phyConsiderations}. This is similar to what a normal person feels just after stepping out of a dark room on a sunny day. % Proper care must be taken while displaying the virtual overlays on the camera feed of physical world when application needs to be used in different lighting conditions.

%[x7]
%Be mindful of sudden changes in brightness. Given the proximity of the screen to the user’s eyes, transitioning the user from a dark scene to a bright scene may cause discomfort as they acclimate to the new level of brightness. It is similar to stepping out of a dark room into the sun.
%	
%
%\subsubsection{Constant Velocity}
%\label{sec:const-vel}
%
%In real life, we feel acceleration and deceleration, but we do not feel velocity.  For example, when flying in an airplane, we feel the takeoff and landing. However, while the plane is traveling at a constant cruising speed of 500mph, we don’t feel anything (assuming there is no turbulence). Similarly, you won’t feel a constant velocity when traveling in a car, only changes in the car’s velocity.
%
%When the user virtually accelerates or decelerates inside of your application, they will not feel the change in real life. The disparity between what they are seeing and what they are feeling may cause discomfort. You can reduce this discomfort by trying to keep the user at a constant velocity when they are moving inside of your app.  

\subsubsection{Other Issues}
\label{sec:other-issues}

Apart from above mentioned limitations, there are some other subjective issues with Google Cardboard such as (a) user discomfort as it pokes on nose, (b) Subjects with spectacles (eye correction) can't wear Google Cardboard, (c) Eye strain when used for longer duration, (d) Claustrophobic, and (e) sometimes causes dizzy feeling if the overlaid virtual content accelerates or decelerates inside the application.

\section{Experiments and Results}
\label{sec:exps}

The experiments were carried out to find out the feasibility of Google Cardboard for AR. 20 engineers and research staff from an industry research lab were selected as participants, comprising 10 male and 10 female, and the ages spanned from 22 to 35 with average 26 years. Their proficiency level was novice with respect to usage of Google Cardboard. Subjects were tasked with conducting an inspection of a Desktop computer using Google Cardboard application in two different settings: (i) quiet reading room, and (ii) work place: (not too quiet and not too loud). The inspection consisted 20 questions related to desktop computer; and the questions are displayed in a sequential manner where the user has to answer them by giving speech input (basically say, either YES or NO or SKIP) once the inspection of particular part mentioned in the question is completed\footnote{Google Cardboard VR SDK for Android is used for developing Android based application for inspection. Google's Speech Recognizer API is integrated for speech-to-text based interaction. Nexus 5 mobile is used in conducting experiments}.

\begin{figure*}[ht!]
\begin{center}
   \includegraphics[width=1\linewidth]{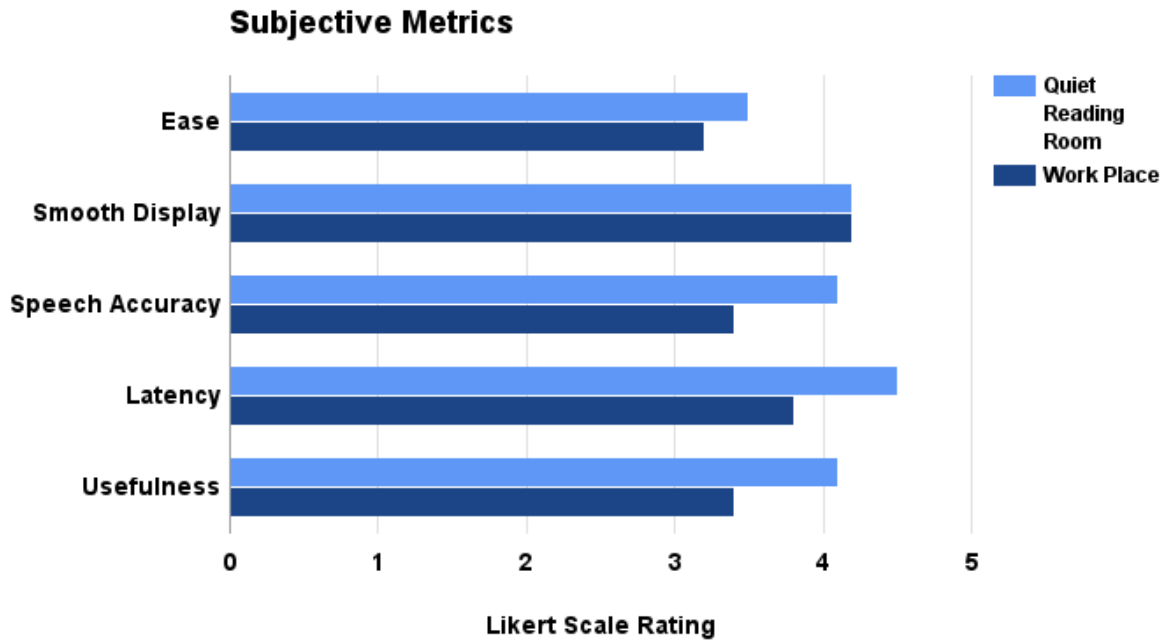}% z1.eps for both charts
\end{center}
   \caption{Mean Likert ratings of  the subjective metrics, namely, (1) Ease, (2) Smooth Display, (3) Speech interface accuracy, (4) Latency and (5) Usefulness over 20 novice subjects for the inspection of Desktop computer. We report two sets of results in a \textit{quiet-reading-room} and \textit{work-place} cases}
\label{fig:exps}
\end{figure*}

A set of subjective metrics were obtained that measure both usability and user experience. These indicators measure human performance and user satisfaction. Users ratings were collected using a five-point Likert scale \ref{robbins2011plotting} ranging from 1 to 5 (1 - Very Poor, 2 - Poor, 3 - Fair, 4 - Good, 5 - Very Good). The Likert scale is commonly used in surveys as it allows the subjects to quantify opinion based items \ref{robbins2011plotting}.

Figure \ref{fig:exps} depicts user ratings of 5 subjective metrics. Most important one being, the usefulness of the method when used for (i) complex inspection process involving huge number of check-lists, and (ii)  maintenance or assembly where user manuals have to be referred many times by the novice users performing tasks. The ease referred to the user comfort and user-friendly interaction that reduced the stress and mental load while carrying out an inspection process. The results obtained conform to the requirements of statistical significance of data obtained by subjective metrics. 

The 5 subjective metrics are as follows:

\begin{itemize}
 
\item[1] \textit{Ease}: How easy was it to perform inspection? Rate the trouble experienced wearing/using the device?
\item[2] \textit{Smooth Display}: Was there a lag in display of Virtual content?
\item[3]\textit{Speech Accuracy}: Was there trouble with speech interaction input method ?
\item[4] \textit{Latency}: Was the system responsiveness for speech inputs user-friendly ?
\item[5] \textit{Usefulness}: Rate the usefulness of application for similar AR maintenance/assembly tasks ? 
\end{itemize}

From the Figure \ref{fig:exps}, it is evident that for \textit{quiet-reading-room} inspection scenario, we can notice high fidelity of \textit{speech-to-text} as there is no surrounding noise in \textit{reading room} setting. For the case of\textit{ work-place}, \textit{speech-to-text} fidelity is relatively poor as there is higher chance of recognizer picking the noise and matching the text output with the pre-defined set of speech keywords. This resulted in higher ratings for \textit{ease} and \textit{usefulness} of the application for \textit{quiet-reading-room} inspection case. 
%
%\begin{table}[ht!]
%\caption{Mean Likert scale ratings of subjective metrics}
%\label{tab:subjMetrics}
%\begin{center}
%\begin{tabular}{ll}
%\toprule
%Parameter & Rating\\
%\midrule
%Usefulness  & 4 \\
%Ease & 3.1  \\
%Smooth Display & 3.2 \\
%Interaction Input &  3.1 \\
%Easy to Wear/Use ? & 3.6
%\bottomrule
%\end{tabular}
%\end{center}
%\end{table}

\section{Discussion}
\label{sec:discussion}

If the limitations are overcome, vision of Cardboard's wider footprint in AR market in next few years is hoped to come true. With the present state of Cardboard design and ability of smartphone, Cardboard can be a good solution for VR applications and few other short duration indoor AR applications. As the speech based  user interaction can not perform well in noisy environments and head rotation is not a convenient user interaction to be used when developing applications for manufacturing industries, we discourage Cardboard for outdoor activity applications with the current state. Choice of optical-see-through AR devices (Google Glass, Vuzix Glass, Daqri Smart Helmet, Microsoft Hololens etc.) will always be preferable for outdoor industry application developments because of their efficient hardware setup with multiple sensors and sophisticated user interaction technologies. 

For aforementioned reasons despite of being an economically viable solution, it posed difficulties to the developers to make decision whether or not to choose Google Cardboard for particular application. Developers are suggested to make an application that works both in (i) stereoscopic view for the use with Cardboard, and (ii) monocular view with additional overlay of on-screen selection buttons for the hand-held or tablet based use. It is always better to provide an option to the user to choose which mode suits. Tablet or smartphone hand-held mode can be preferable in few use cases for the following reasons: (a) solves many non-technical problems related to Google Cardboard as user is always free to interact with objects without relying completely on on-screen camera feed, (b) camera feed is necessary only when there is a need for virtual objects integration with physical world objects, and (c) speech recognition overhead will not be there as it uses overlay of on-screen selection buttons for user interaction saving the battery power. The only disadvantage using hand-held mode is that it is difficult to use for maintenance tasks where user has to work with both hands.

\section{Future Opportunities}
\label{sec:futOpp}

Google Cardboard design requires more attention for addressing the fundamental issues before it becomes widely accepted in AR market. And also, smartphones should get more powerful with better displays, camera, and processors; and manufacturers working on designing headsets similar to Cardboard should provide better mechanism with additional modifications such as (a) taking off side flaps to widen FoV, (b) placing cushion for nose, and (c) slide mechanism to adjust the inter lens distance and distance between lens and smartphone.

With the current state of smartphone technology, Google Cardboard based AR applications can be effectively used in areas such as travel, e-commerce, 3D gaming and education. Qualcomm, with its pioneering work in mobile AR, encourages AR applications for Cardboard thorugh Vuforia SDK providing side-by-side camera feed with 3D calibration and lens distortion correction \ref{vuforia}. Vuforia overlays VR content by detecting and tracking a \emph{Image Target}. Common uses of \emph{Image Targets}  include recognizing and augmenting printed media and product packaging for marketing campaigns, gaming and visualizing products in the environment where the product was intended to be used \ref{img_target}. Vuforia's markerless recognition is limited; this helps in applications such as augmenting 3D content on toys or instructional manuals overlaid on consumer products.

Cardboard based AR can be essential guide in providing educational resources to re-create historical events, activate conventional books into 3D graphics by recognizing \emph{Image Targets}. It can also serve as a visual aid for training novice users in manufacture industry in constrained environment.

\section{Conclusion}
\label{sec:Concln}

Google Cardboard has drawn tremendous attention as its economically viable facilitating wider reach among both developers and users. In this paper, we provided a comprehensive overview of Google Cardboard design and its evolution for VR and AR applications. We also discussed stereo rendering video-see-through based AR applications development for Google Cardboard by highlighting the technical limitations and subjective considerations from users. Our study is a guide for Android/iOS developers as there are no published scholarly articles or well documented studies with user/developer's experience. If technical and non-technical limitations are overcome, vision of Cardboard's wider footprint in AR market in next few years is hoped to come true. With the current state of smartphone technology and existing design of Google Cardboard, we suggested impactful future opportunities for Google Cardboard based AR applications.

\bibliography{scifile}

\section{References}

\begin{enumerate}

\item \label{speech} ANDROID SPEECH API. Android Speech Recognizer API. URL: http://developer.android\\.com/reference/android/speech/
SpeechRecognizer.html. Accessed: 2016-04-22.
\item \label{branstetter}
BRANSTETTER , B. Cardboard is everything Google Glass
never was. URL: http://kernelm\\ag.dailydot.com/issue-sections/
staff-editorials/13490/google-cardboard-review-plus.
Accessed: 2016-04-22.
\item \label{google_blog}
CARDBOARD JOURNEY. Google Official Blog.
URL: https://googleblog.blogspot.in/2016\\/01/unfolding-virtual-journey-cardboard.html. Accessed: 2016-06-29.
\item \label{cardboard_sdk}
CARDBOARD SDK. Google Cardboard SDK Overview. URL: https://developers.google.com\\/cardboard/overview. Accessed: 2016-
04-22.
\item \label{phyConsiderations}
CARDBOARD SDK. Physiological Consideration in Designing for
Google Cardboard. URL: https://www.google.com/design/spec-vr/
designing-for-google-cardboard/physiologi\\cal-considerations.
html\#. Accessed: 2016-04-22.
\item \label{opengl}
OPENGL API. Open GL Graphics API for Android. URL: http://
developer.android.com/gu\\ide/topics/graphics/opengl.html. Ac-
cessed: 2016-04-22.
\item \label{pierce}
PIERCE , D. Google Cardboard Is VR's Gateway Drug. URL: http:
//www.wired.com/\\2015/05/try-google-cardboard/. Accessed:
2016-04-22.
\item \label{robbins2011plotting}
ROBBINS , N. B., AND HEIBERGER , R. M. 2011. Plotting likert
and other rating scales. In Proceedings of the 2011 Joint Statistical Meeting, 1058-1066.
\item \label{shanklin}
SHANKLIN , W. Google Cardboard 2: Hands-on. URL: http://www.
gizmag.com/google-cardboard-2-review-initial/37777/. Ac-
cessed: 2016-04-22.
\item \label{sutherland}
SUTHERLAND , I. E. 1968. A head-mounted three dimensional
display. In Proceedings of the December 9-11, 1968, fall joint
computer conference, part I, ACM, 757-764.
\item \label{technavio}
TECHNAVIO. The Pros and Cons of Google
Cardboard. URL: http://www.technavio.com\\/blog/
the-pros-and-cons-of-google-cardboard\#sthash.o2YVHp0h.
dpuf. Accessed: 2016-04-22.
\item \label{unity}
UNITY SDK. Unity SDK Reference for Google Cardboard. URL: https:
//developers.google.com/cardboard/unity/reference. Accessed:
2016-04-22.
\item \label{van}
VAN KREVELEN , D., AND POELMAN , R. 2010. A survey of
augmented reality technologies, applications and limitations. In-
ternational Journal of Virtual Reality 9, 2, 1.
\item \label{hypergrid}
VR HEADSETS. Hypergrid Business:
Virtual Reality Headsets. URL: http://www.hypergrid\\business.com/faq/
best-virtual-reality-headsets/. Accessed: 2016-04-22.
\item \label{vr_profile}
VR PROFILE GENERATOR. Google Viewer Profile
Generator. URL: http://www.google.c\\om/get/cardboard/
viewerprofilegenerator/. Accessed: 2016-04-22.
\item \label{vuforia} VUFORIA SDK. Qualcomm's Vuforia SDK: Developing for
Google Cardboard. URL: https://developer.vuforia.com/library/
articles/Solution/Developing-for-Google-Cardboard. Accessed: 2016-04-22.
\item \label{img_target}
VUFORIA SDK. Qualcomm's Vuforia SDK: Image Target based
Recognition. URL: https://developer.vuforia.com/library/articles/
Training/Image-Target-Guide. Accessed: 2016-04-22.
\item \label{perla_inspection}
R Perla, R Hebbalaguppe, G Gupta,
G Sharma, E Hassan, M Sharma, L Vig, and G Shroff. 2016. An AR Inspection Framework:
Feasibility Study with Multiple AR Devices. In International
Symposium on Mixed and Augmented Reality. IEEE, 2016.
\item \label{hegde_gestures}
S Hegde, R Perla, R Hebbalaguppe, and E Hassan. 2016. GestAR: Real Time Gesture Interaction
for AR with Egocentric View. In International Symposium on Mixed and Augmented Reality. IEEE, 2016.
\item \label{mohatta_gestures}
S. Mohatta, R. Perla, G. Gupta, E. Hassan, and R. Hebbalaguppe. Robust hand gestural interaction for smartphone based ar/vr applications. In Winter Applications on Computer Vision. IEEE, 2017.
\item \label{gupta_roi}
A Gupta, S Mohatta, J Mourya, R Perla, E Hassan, R Hebbalaguppe. Hand Gesture based Region Marking for Tele-support using Wearables. In Embedded Vision workshop, CVPR, IEEE, 2017.
\end{enumerate}

% Following is a new environment, {scilastnote}, that's defined in the
% preamble and that allows authors to add a reference at the end of the
% list that's not signaled in the text; such references are used in
% *Science* for acknowledgments of funding, help, etc.
%
%\begin{scilastnote}
%\item We've included in the template file \texttt{scifile.tex} a new
%environment, \texttt{\{scilastnote\}}, that generates a numbered final
%citation without a corresponding signal in the text.  This environment
%can be used to generate a final numbered reference containing
%acknowledgments, sources of funding, and the like, per {\it Science\/}
%style.
%\end{scilastnote}

%
%
%\clearpage
%
%\noindent {\bf Fig. 1.} Please do not use figure environments to set
%up your figures in the final (post-peer-review) draft, do not include graphics in your
%source code, and do not cite figures in the text using \LaTeX\
%\verb+\ref+ commands.  Instead, simply refer to the figure numbers in
%the text per {\it Science\/} style, and include the list of captions at
%the end of the document, coded as ordinary paragraphs as shown in the
%\texttt{scifile.tex} template file.  Your actual figure files should
%be submitted separately.
%

\end{document}